\documentclass[11pt,onecolumn,english,amsart,preprintnumbers,amsmath,amssymb,floatfix,nofootinbib]{revtex4}
\usepackage{tikz,xcolor}
\usepackage[colorlinks = true,
linkcolor = red,
urlcolor  = blue,
citecolor = red,
anchorcolor = blue]{hyperref}

\definecolor{lime}{HTML}{A6CE39}
\DeclareRobustCommand{\orcidicon}{%
	\begin{tikzpicture}
	\draw[lime, fill=lime] (0,0)
	circle [radius=0.16]
	node[white] {{\fontfamily{qag}\selectfont \tiny ID}};
	\draw[white, fill=white] (-0.0625,0.095)
	circle [radius=0.007];
	\end{tikzpicture}
	\hspace{-2mm}
}

\foreach \x in {A, ..., Z}{%
	\expandafter\xdef\csname orcid\x\endcsname{\noexpand\href{https://orcid.org/\csname orcidauthor\x\endcsname}{\noexpand\orcidicon}}
}

\usepackage[toc,page]{appendix}
\usepackage{epsfig}
\usepackage{graphics}
\usepackage{latexsym}
\usepackage{amsmath}
\usepackage{amssymb}
\usepackage{rotating}
\usepackage{subfigure}
\usepackage{bm}
\usepackage{color}
\usepackage{ulem}
\usepackage{booktabs}
\usepackage{caption}
\usepackage{float}
\usepackage{multirow}

\usepackage{placeins}
\usepackage{needspace}
\usepackage{xcolor}
\definecolor{color}{RGB}{0,0,0} 
\usepackage{siunitx}

\newcommand{\RNum}[1]{\uppercase\expandafter{\romannumeral #1\relax}}

\begin{document}


\title { A comparative study of different approaches for heavy quark energy loss, based on the latest experimental data }

\author{Marjan Rahimi Nezhad$^{1}$\orcidA{}}
\email{marjanrahimi29@gmail.com}

\author{Fatemeh Taghavi-Shahri$^{1}$\orcidA{}}
\email{taghavishahri@um.ac.ir (Corresponding author)} 

\author{Sharareh Mehrabi Pari$^{1}$\orcidC{}}
\email{sharareh.mehrabi.pari@gmail.com}

\author{Kurosh Javidan$^{1}$\orcidC{}}
\email{javidan@um.ac.ir}

\affiliation {
$^{(1)}$Department of Physics, Ferdowsi University of Mashhad, P.O.Box 1436, Mashhad, Iran  }
\date{\today}

%

\begin{abstract}\label{abstract}
This paper \textcolor{color}{examines collisional and radiative energy loss of heavy quarks in Quark-Gluon Plasma and} presents a comparative analysis of three different methods for energy dissipation.
The study focuses on calculation of the nuclear modification factor ($R_{AA}$) of charm quarks in Pb-Pb collisions at $\sqrt{S_{NN}}=$ \SI{5.02}{\tera\electronvolt}. 
All three methods are examined using the same numerical evolution based on the well-known Fokker-Planck equation by considering critical phenomena like a non-equilibrium state at the onset of heavy ion collisions. The outcomes of each approach are compared with the latest data from ALICE and ATLAS experiments spanning from 2018 to 2022. This study aims to compare the degree of agreement between each approach and recently obtained experimental data, in the intermediate and high $P_T$ regions.\\
\end{abstract}


\maketitle

\tableofcontents{}

%
\section{Introduction}\label{sec:Introduction}
%
Interaction between quarks and gluons is described by the Quantum Chromo Dynamics (QCD) theory, in which quarks act as constituents of hadrons, and gluons act as quantum bosons \cite{Seymour:2005hs}. Two prominent features of QCD are asymptotic freedom and confinement. Asymptotic freedom means that the interaction between quarks is weak when they are close to each other, but as the quarks move away from each other, this force becomes stronger and increases, which is called confinement. As a result of the asymptotic freedom, when matter reaches extremely high temperatures and/or densities, the strong interaction weakens, and quarks and gluons are freed from each other. In other words, at high temperatures and/or densities, hadrons will melt and the degrees of freedom of matter will be quarks and gluons. In this state, a fluid called Quark-Gluon Plasma (QGP) is formed \cite{QGP.Collins:1974ky, QGP.Shuryak:1977ut, QGP.Martinez:2013xka}.\\
Studies indicate that matter was in the plasma phase until a few microseconds after the Big Bang, and then a phase transition occurred. Therefore, studying the transition of the quark phase to the hadron phase and investigating the quark-gluon plasma properties are very important in understanding the evolution of the early universe. Furthermore, matter in heavy cosmic bodies like neutron stars may exist in this state due to its extremely dense composition. Hence, this is a significant issue in Astrophysics as well \cite{Hagedorn:1965st}.\\
QGP was first theorized in the 1970s, but experiments at the Relativistic Heavy Ion Collider (RHIC) and later at the Large Hadron Collider (LHC) confirmed the existence of QGP in the late 1990s. These experiments involve colliding heavy ions, such as gold or lead, at very high energies, creating a hot and dense environment that allows quarks and gluons to move freely and interact strongly with each other  \cite{STAR:2005gfr, PHENIX:2004vcz}. This highly excited state of matter, whose main constituents are light quarks and gluons, displays properties similar to a nearly perfect fluid and can be successfully described by hydrodynamic models \cite{Ollitrault:2007du, Song:2010mg, Becattini:2014rea}. Two different conditions are needed to describe the QGP by hydrodynamics; the first is that the system should have a local thermal equilibrium for a sufficient period of time, and the second one is that the scale of interactions (or mean free distance of particles) should be much smaller than the dimensions of the system.
Hydrodynamics could be considered a macroscopic effective field theory that has the ability to investigate the evolution of non-equilibrium systems.\\
Heavy quarks such as b and c quarks play an essential role in studying the properties of quark-gluon plasma created in heavy-ion collisions \cite{HQ.vanHees:2005wb, HQ.Rapp:2009my}. These quarks are formed in the early stages of the collision. Due to their large mass, they reach equilibrium with the environment later and may even leave the plasma without reaching equilibrium. So they are good witnesses to the whole space-time history of the deconfined medium. In order to study the evolution of heavy quarks in QGP, a possible approach is to examine the time evolution of their distribution function in the transverse momentum plane. It is reasonable to assume that these heavy particles in a non-equilibrium state undergo Brownian motion within a heat bath that is in thermodynamic equilibrium. The Fokker-Planck equation can be used to obtain the temporal evolution of the transverse momentum spectrum of heavy quarks. When heavy quarks pass through the plasma, they interact with the QGP constituents and lose energy through radiation and elastic collisions. The energy loss of these quarks provides information about the properties of the QGP, such as its temperature and viscosity. In addition, the study of heavy quark energy loss is important for understanding the mechanism of jet quenching, which is the suppression of high-energy partons in the QGP.\\
In this article, we are going to investigate the evolution of the charm quark distribution function in Pb-Pb collision at $\sqrt{S_{NN}}=$ \SI{5.02}{\tera\electronvolt}. In our evolution process, we consider different approaches for collisional energy loss, along with radiation energy loss. Eventually, by calculating the nuclear modification factor, $R_{AA}$, we are able to compare theoretical results with the most recent experimental data from LHC, in order to determine which method of energy dissipation is most compatible with new experimental data. This paper is organized as follows: In section (\ref{sec: Methods}), we review the Fokker-Planck equation and the evolution of the QGP system. We also introduce different methods of energy loss that we have examined. Section (\ref{sec: Results}) is focused on calculating the nuclear modification factor and presenting our theoretical results. Our $R_{AA}$ results for each energy loss model are compared with new data from ALICE and ATLAS. Finally, the conclusion is given in section (\ref{sec: conclusion}). 
%
\section{Methods}\label{sec: Methods} 
In this section, we describe details of our modeling framework. To begin with, it would be helpful to review the stages of quark-gluon plasma formation. Heavy-ion collisions pass through various stages, from collision to hadronization. The collision initially produces a fireball of quarks and gluons known as quark-gluon plasma. After a while, the system quickly reaches local thermodynamic equilibrium, and high-energy partons lose energy through passing the plasma. As the system continues to expand and cool, all interactions stop and the system reaches the freeze-out temperature ($T_f$). In this state, dynamic information remains constant and hadrons are formed.\\
\subsection{ System evolution }
To study the QGP system, various models could be used to calculate the time evolution of dynamic parameters such as temperature and viscosity. Here we consider the time dependence of temperature as follows \cite{Chattopadhyay:2018apf, Grozdanov:2015kqa}:
\begin{equation}\label{Eq:eq1}
T(\tau) = T_0\left(\frac{\tau_0}{\tau}\right)^{\frac{1}{3}}\left[1+\frac{2}{3\tau_0T_0}\frac{\eta}{s}\left(1-\left(\frac{\tau_0}{\tau}\right)^{\frac{2}{3}}\right)\right]
\end{equation}
where $T_0$ and $\tau_0$ are initial temperature and proper time respectively, and $\frac{\eta}{s}$ is viscosity to entropy ratio, \textcolor{color}{which has been considered as a constant with a value of $1/4\pi$ \cite{Ruggieri:2013ova}.
The initial proper time has been taken as $\tau_0=$ \SI{0.33}{\femto\meter\per\si{\second}} and the initial temperature has been set to $T_0=$ \SI{403}{\mega\electronvolt}. For the hydrodynamics equations, see section C in the appendix}. \\
Also, using a temperature-dependent function for the running coupling $ \alpha_s(T)$ is essential because the temperature is a critical scale that controls the QCD coupling in the QGP system \cite{Braaten:1989kk}: 
\begin{equation}
	\alpha_s(T) = \frac{6\pi}{(33-2N_f)\ln\left(\frac{19T}{\Lambda_{MS}}\right)}
\end{equation}
where we assume $N_f=3$ as the number of active flavors in the QGP and the QCD cut-off parameter has been taken as $\Lambda_{MS}=$ \SI{80}{\mega\electronvolt}.\\
The heavy quarks evolution in the QGP system can be described by different approaches. 
In this article, we will employ the Fokker-Planck equation which is a simplified form of the Boltzmann equation, in the case that the momentum transform is small \cite{Dong:2019unq, Zhao:2020jqu}. The Fokker-Planck equation provides a suitable framework for investigating the temporal evolution of heavy quarks \cite{FP.Dong:2019byy}. This equation was first introduced by Fokker and Planck to explain the Brownian motion of particles in a fluid. According to this equation, the temporal evolution of the distribution function of heavy quarks is given by:
\begin{equation}
\frac{\partial}{\partial t}f(p,t) = -\frac{\partial}{\partial p_{i}}\left[A_{i}(p)f(p,t)\right] + \frac{\partial^2}{\partial p_{i}\partial p_{j}}\left[D_{ij}(p)f(p,t)\right].
\end{equation}
To solve this equation, we require three input parameters: \\
The initial distribution function of heavy quarks ($f_{in}(p,t)$), drag ($A_{i}(p)$) and diffusion ($D_{ij}(p)$) coefficients. \textcolor{color}{The i and j indices denote different spatial directions, which can be eliminated since we consider a spatially uniform QGP}.\\
The drag and diffusion coefficients are determined by the non-equilibrium energy dissipation of particles in a thermal environment. 
The energy dissipation of heavy quarks in a plasma environment occurs through two main processes:
(1) collisions with other particles and (2) gluon bremsstrahlung or radiation due to interactions of heavy quarks with other quarks, anti-quarks, and gluons present in the thermal bath. \\
Therefore, the drag coefficient can be obtained using the following relation \cite{Diffusion.Das:2010tj}: 
\begin{equation}
A(p) \propto -\frac{1}{p}\frac{dE}{dL}
\end{equation}
while we consider energy loss in both cases:
\begin{equation}
	\frac{dE}{dL} = (\frac{dE}{dL})_{coll}+ k (\frac{dE}{dL})_{rad}
\end{equation}
The value of $k$ is uncertain and needs to be determined through an optimization process. \textcolor{color}{This value shows the impact of radiation term in total energy loss.}\\
The diffusion coefficient can be determined using Einstein's relation when there is a weak coupling between the heavy quarks and the thermal bath \cite{Diffusion.Srivastava:2016igg,Diffusion.Das:2010tj,Diffusion.Akamatsu:2008ge}:
\begin{equation}
D(p) = T A(p) E
\end{equation}
\textit{T} represents temperature of the thermal bath and \textit{E} represents the energy of heavy quarks.\\
Note that the drag coefficient carries information about dynamics of heavy quark collisions with the medium and is expected to be determined by the properties of the thermal bath. Therefore, the most critical point for finding the time evolution of the HQ distribution function is calculating the drag force acting on the HQ or the corresponding rate of energy loss per unit distance of the HQ path in QGP. One should use a gauge invariant field theory that does not have any infrared divergence to properly account for thermal effects and obtain accurate outcomes for these values. By calculating energy loss, we have all the parameters to solve the FP equation and study HQ evolution from the onset of plasma formation to reaching the critical temperature and hadronization.\\
\\ 
%
\subsection{Energy loss approaches}
The asymptotic freedom of QCD implies that, for a quark-gluon plasma at a sufficiently high temperature, the rate of energy loss $dE/dx$ can be calculated using perturbation theory based on the running coupling constant $\alpha_s(T)$. Unfortunately, it is not possible to compute $dE/dx$ directly by evaluating the tree-level Feynman diagrams for scattering off of thermal quarks and gluons in the plasma. There are different divergences due to the long-range interactions mediated by the gluon. Indeed, gluon exchange diagrams give rise to logarithmically infrared divergent integrals over the momentum transfer \textit{q} of the gluon.\\
In this study, we compare three different approaches to calculate collisional energy loss for heavy quarks in the QGP. Each of these approaches has addressed the divergence problem in its own way. The Fokker-Planck equation is employed to investigate each approach by evaluating the distribution function of HQs from the time of plasma equilibration to hadronization. The $R_{AA}$ plot is utilized to compare the degree of agreement of each approach with the latest experimental data.\\
The first approach for collisional energy loss (Model A) has been calculated by Bjorken \cite{Bjorken:1982tu}. It is indeed the first calculation of the heavy quark energy loss due to QGP-HQ interaction. He calculated the energy loss of a massless quark due to elastic scattering off of the QGP constituents by averaging the cross section multiplied by the mean energy transfer over the thermal distribution. \textcolor{color}{This approach regulates infrared singularity by proposing a Debye mass derived from the Quantum Electrodynamics (QED) framework.}\\ 
The second approach (Model B), proposed by Thoma and Gyulassy \cite{Thoma:1990fm}, combines techniques of plasma physics with high-temperature QCD \cite{Pisarski:1989cs} in order to calculate collisional energy loss. Through this method, $dE/dx$ is computed using the induced chromoelectric field in the wake of a high-energy quark. That reduced field is related to the longitudinal and transverse dielectric functions, which in turn, can be expressed in terms of the gluon self-energy. An advantage of this approach is its ability to regulate infrared singularity through the Debye mass automatically, \textcolor{color}{ as well as ultraviolet singularity by introducing $K_{max}$}. \\
The last approach studied in this research (Model C) is proposed by Braaten and Thoma \cite{Braaten:1991we}, which includes calculating the energy loss of a quark with energy E in two different limits: $E \ll \frac{M^2}{T} \text{ and } E \gg \frac{M^2}{T}$. In this method, soft and hard contributions to the energy loss are calculated separately and added together. This approach utilizes the hard-thermal loop (HTL) framework \cite{Braaten:1989kk, Braaten:1991we}.\\
The radiative energy loss of heavy quarks has been calculated using the proposed model in Ref. \cite{Saraswat:2017vuy}. This formalism has been constructed by considering the reaction operator formalism (DGLV) and employing the generalized dead cone approach \cite{Saraswat:2015ena, Abir:2012pu}. The DGLV approach \cite{Gyulassy:2000fs, Djordjevic:2003zk, Wicks:2005gt} relies on expanding the quark energy loss based on the number of scatterings encountered by the quark as it moves through the medium. The single hard scattering limit considers only the leading order term. See the appendix for more information.\\
%
\subsection{Nuclear modification factor}
Quark-gluon plasma formation cannot be directly observed in the laboratory because the formed matter quickly cools down and has a very short lifetime (on the order of $10^{-23}$ seconds). What is observed and recorded by detectors are only photons, leptons, and stable final hadrons. Therefore, we need measurable quantities that are dependent on the characteristics of the initial stages of the system to obtain information about the early stages of plasma formation. \\
Here, we introduce one of the most important signals of plasma formation which is the nuclear modification factor ($R_{AA}$). This quantity represents the ratio of the number of electrons produced from semi-leptonic decay of mesons per unit rapidity and transverse momentum in nuclear-nuclear collisions to the same value in the proton-proton collisions \cite{Raa.Miller:2007ri, CMS:2016xef, Raa.Zigic:2019sth}:
\begin{equation}
	R_{AA}(p_T) = \frac{\left(\frac{dN^e}{dP_T^2 dy}\right)^{A-A}}{N_{coll} \times \left(\frac{dN^e}{dP_T^2 dy}\right)^{p-p}}
\end{equation}
The term "$N_{coll}$" in the denominator represents the number of nucleon-nucleon collisions in nucleus-nucleus collisions and can be estimated via Glauber model calculations \cite{Miller:2007ri}.\\
The nuclear modification factor quantifies the amount of energy loss during nucleus collisions due to heavy quarks' transportation in the partonic medium. When there is no creation of the quark-gluon plasma, the nuclear modification factor is equal to one, which signifies the non-existence of a novel medium.  However, if the value of this factor is less than one, it reveals the interaction between high-energy jets and the thermal environment formed during the collision of energetic nuclei.\\
It should be noted that what is observed in detectors are electrons produced from the decay of D and B mesons. Therefore, to obtain more precise results, the corresponding hadron distribution functions can be derived by utilizing a suitable fragmentation function on the output of the FP equation. However, the application of the fragmentation function to the final result has a negligible impact on our comparison and can be ignored \cite{Tripathy:2017kwb, Tripathy:2016hlg, Qiao:2020yry, vanHees:2005wb}, \textcolor{color}{see appendix (Section B) for more details}. In this work, the partonic distribution functions are directly divided by each other to calculate the nuclear modification factor. Adding a \textcolor{color}{scaled factor (N)} would result in the following outcome:
\begin{equation}
    R_{AA} = \frac{1}{N} \frac{f_f(P_T)^{A-A}}{f_i(P_T)^{P-P}}
\end{equation}
\\
\FloatBarrier
\section{Results and  Discussion} \label{sec: Results}
\begin{figure}[htb]
	\begin{center}
		\vspace{0.4cm}
	\resizebox{0.65\textwidth}{!}{\includegraphics{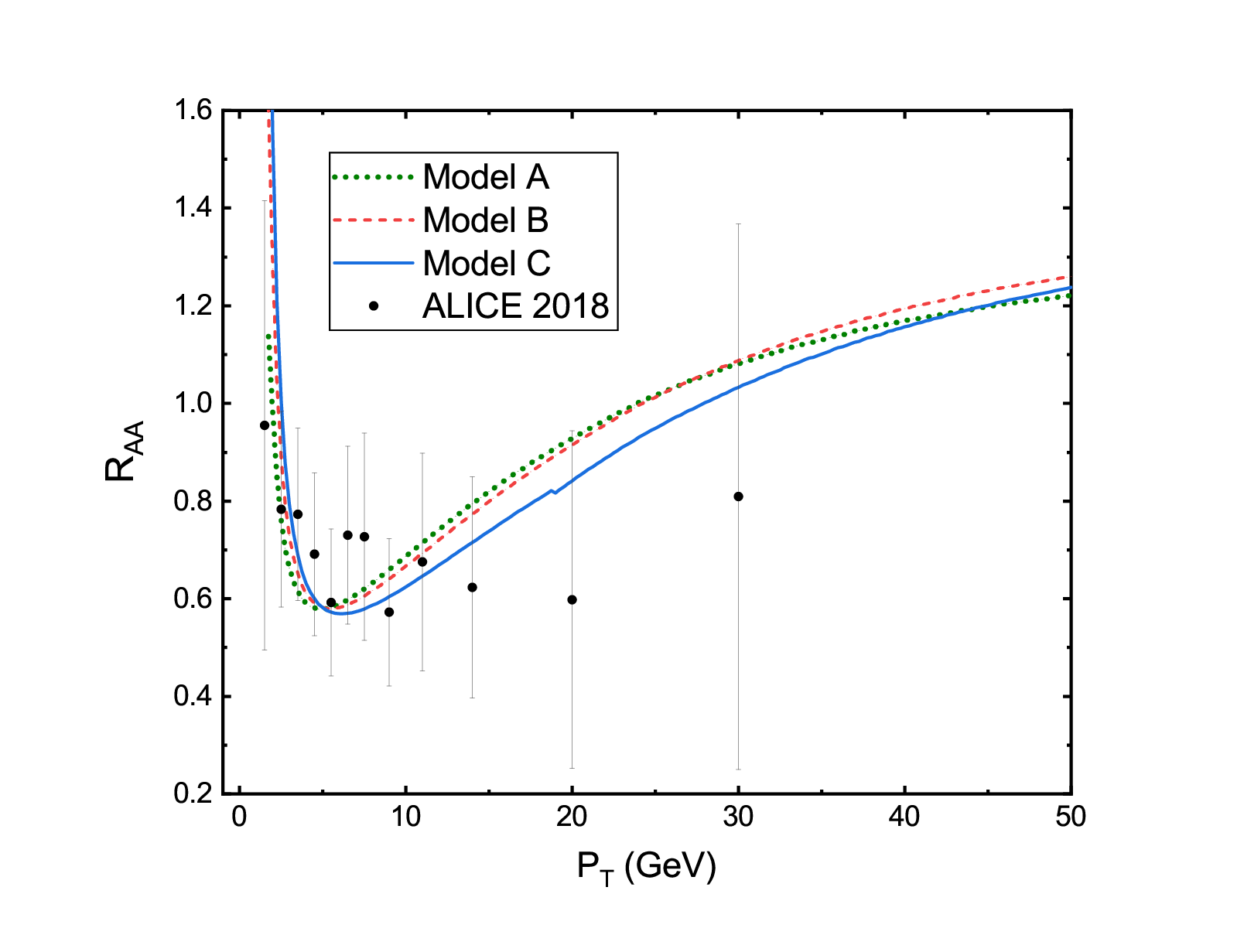}}
	\caption{ Nuclear modification factor of charm quark in the (60–80\%) centrality Pb-Pb collisions at $\sqrt{S_{NN}}=5.02$ TeV compared with ALICE 2018 data at mid-rapidity ($|y| < 0.5$) \cite{ALICE:2018lyv}. }\label{fig: ALICE 2018}
\end{center}
\end{figure}

\begin{figure}[htb]
	\begin{center}
		\vspace{0.4cm}
	\resizebox{0.65\textwidth}{!}{\includegraphics{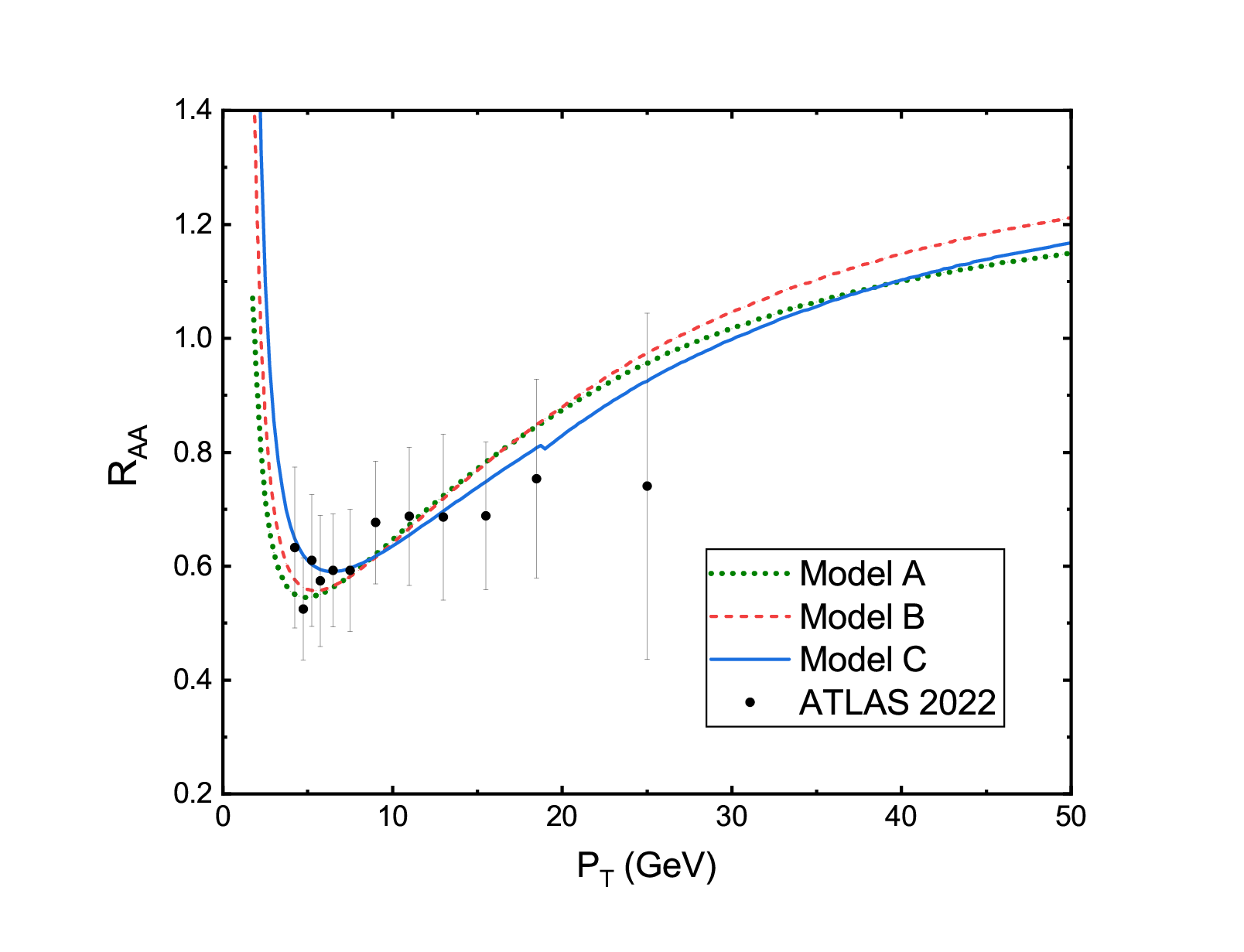}}
	\caption{ Nuclear modification factor of charm quark in the (40–60\%) centrality Pb-Pb collisions at $\sqrt{S_{NN}}=5.02$ TeV compared with ATLAS 2022 data at pseudo-rapidity ($|\eta|<2$) \cite{ATLAS:2021xtw}.}\label{fig: ATLAS 2022}
\end{center}
\end{figure}

\begin{figure}[htb]
	\begin{center}
		\vspace{0.4cm}
	\resizebox{0.45\textwidth}{!}{\includegraphics{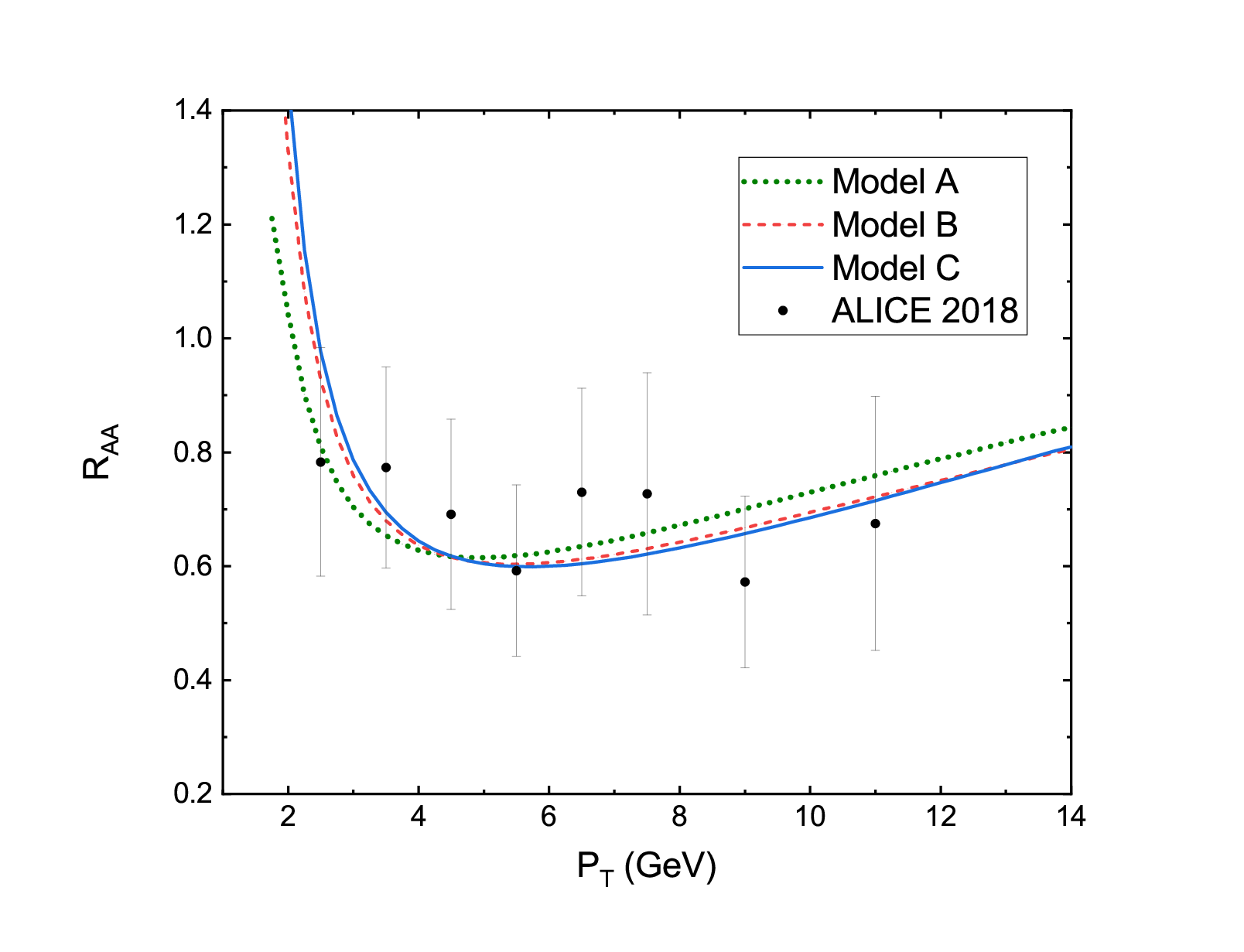}}
        \resizebox{0.45\textwidth}{!}{\includegraphics{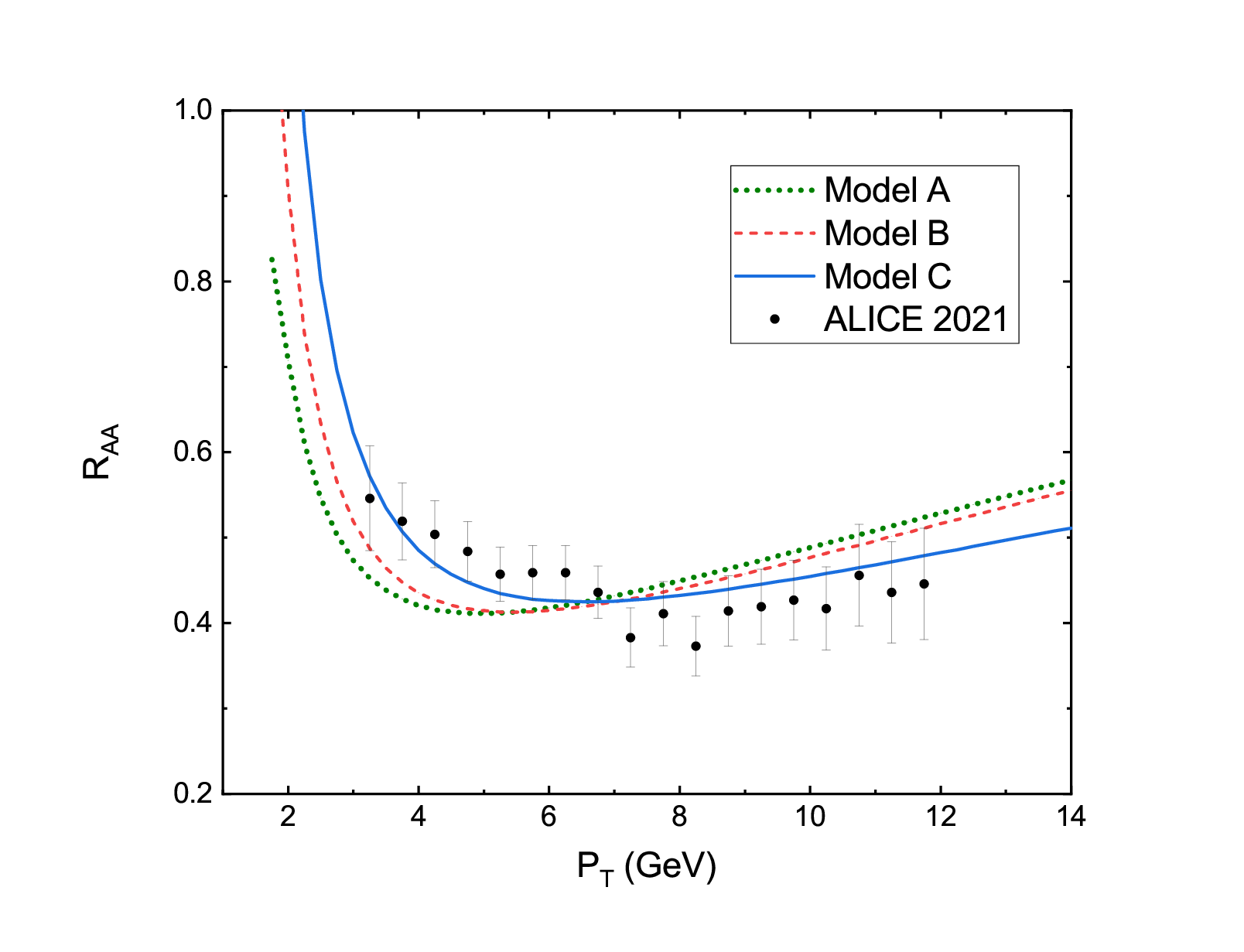}}
        \resizebox{0.45\textwidth}{!}{\includegraphics{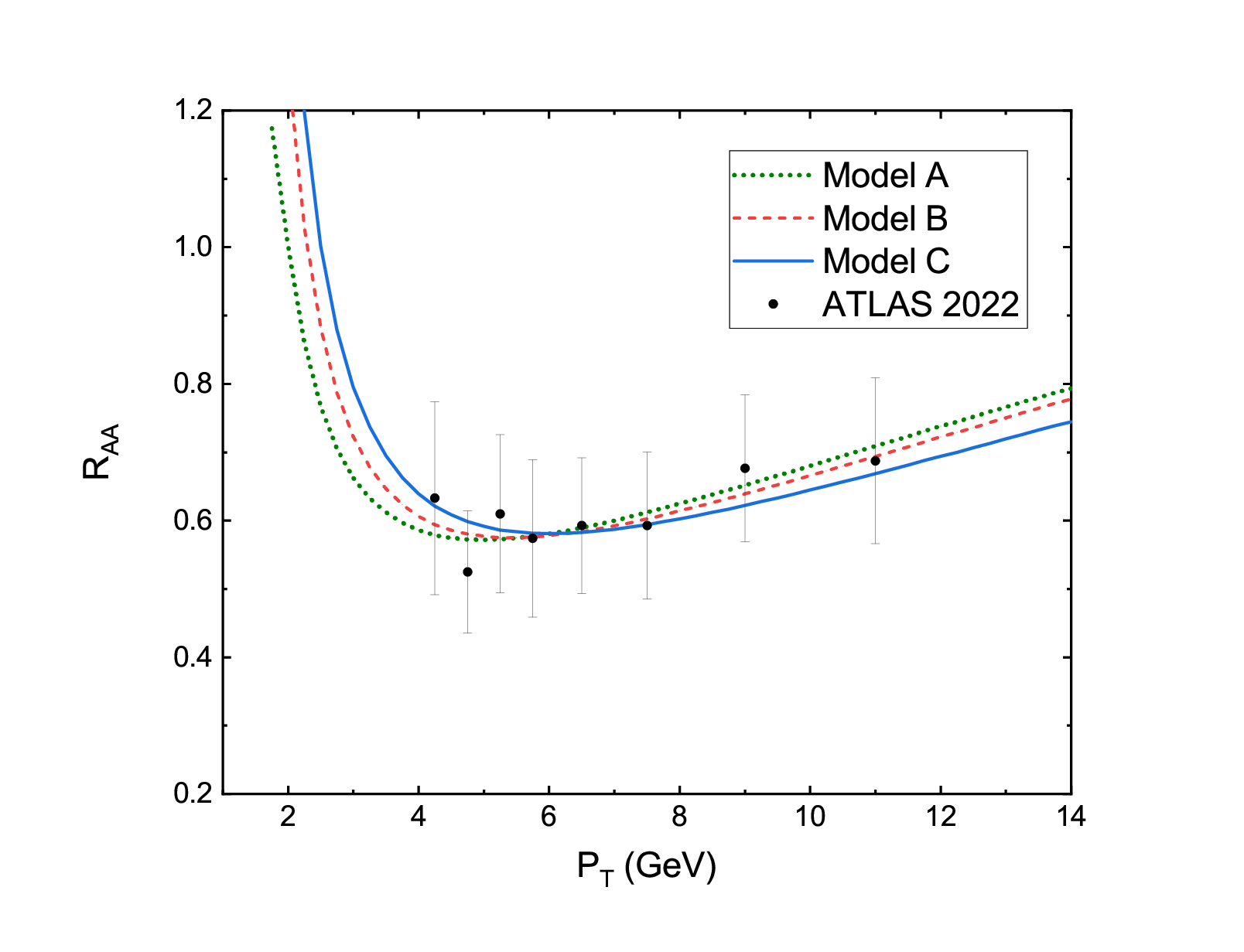}}
       \resizebox{0.45\textwidth}{!}{\includegraphics{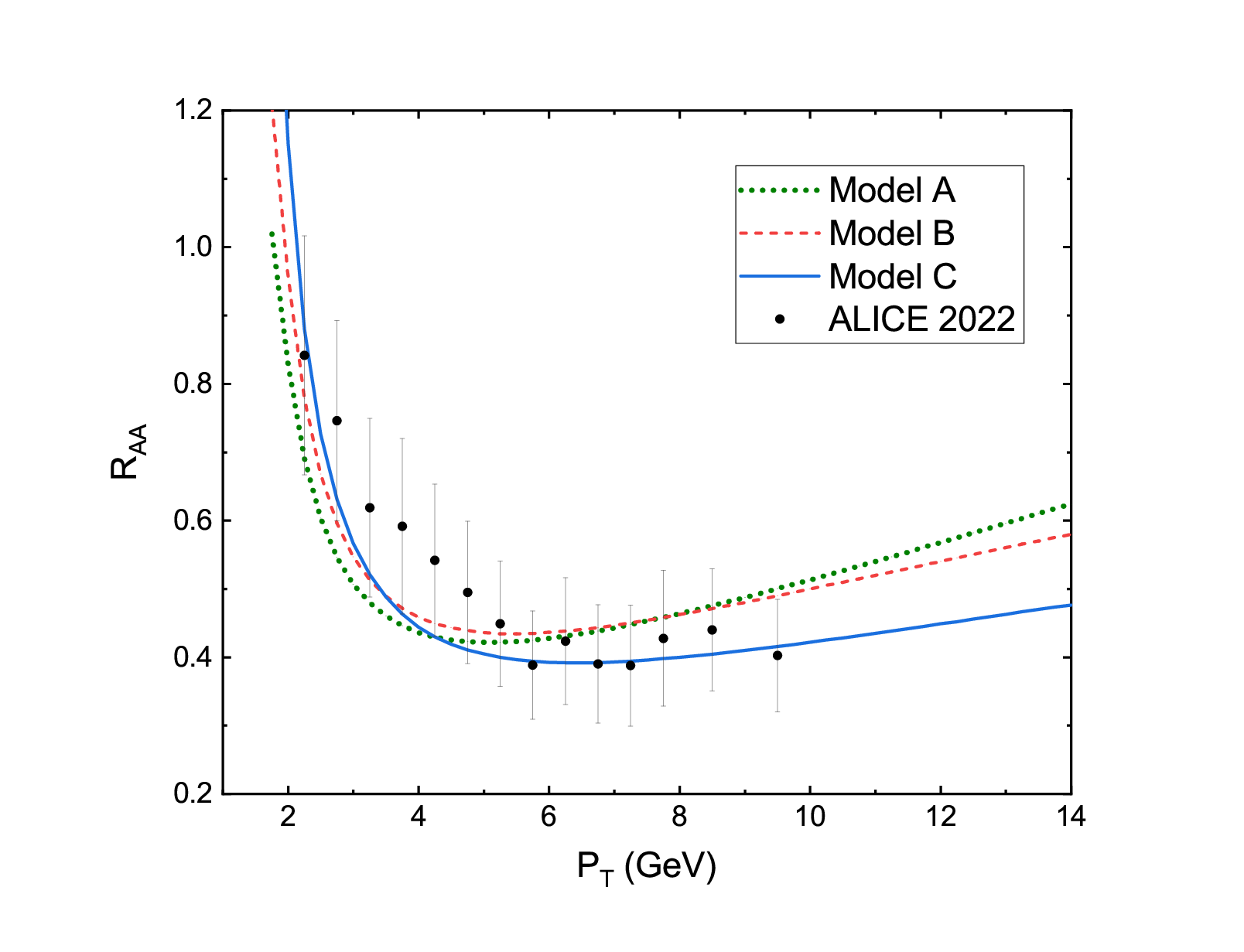}}
	\caption{ Nuclear modification factor
		of charm quark at 5.02 TeV fitted on experimental data for the intermediate range of $P_T$. }\label{fig: Intermediate $P_T$}
\end{center}
\end{figure}

In this section, we calculate the nuclear modification factor of the charm quark in a Pb-Pb collision at a center-of-mass energy of \SI{5.02}{\tera\electronvolt}. The parton distribution functions required to perform this calculation are evaluated using the Fokker-Planck equation. FP equation is solved numerically at third-order relativistic hydrodynamics \cite{Numerical solution} until the fireball cools down to its freeze-out temperature. The evolution of HQ distributions is calculated from initial proper time  $\tau_0=$ \SI{0.33}{\femto\meter\per\si{\second}} to thermal freeze-out $T_c=$ \SI{155}{\mega\electronvolt} for LHC \cite{Freeze-Out Parameters.Chatterjee:2015, Freeze-Out Parameters.HotQCD:2014kol}. The initial transverse momentum distribution of charm quark is obtained from \cite{Sheibani:2021ovo, Modarres:2021gva, Olanj:2020lkt}. \textcolor{color}{In reference \cite{Sheibani:2021ovo}, distribution functions are obtained from the APFEL legacy framework, using the nCTEQ15 model and EMC effect, to obtain the bounded quarks inside the nuclei at a specific center-of-mass energy.
In the next step, the Pythia8 simulation code is used to compute the initial spectra of HQs after the Pb-Pb collision.
Details of converting distribution functions in terms of x-variable to transverse-momentum-dependent PDFs can be found in formula (1) of this article \cite{Sheibani:2021ovo}.} \\
To compute drag and diffusion coefficients, besides considering radiation energy loss, we employ three distinct approaches which previously introduced for evaluating collisional energy loss. The outcomes of each approach are compared with the most recent data from ALICE and ATLAS in 2018, 2021 and 2022 \cite{ATLAS:2021xtw, ALICE:2018lyv, ALICE:2020sjb, ALICE:2021rxa}. Our purpose is to determine which approach is most consistent with the experimental results.\\
The final results are optimized to fit on experimental data by adjusting initial parameters such as k and N, and minimizing the unweighted Chi-squared value:
\begin{equation}
	\chi^2=\sum_i \frac{(R_{AA}^{exp} (P_T(i))-R_{AA}^{th} (P_T(i))^2}{\sigma_i^2}
\end{equation}
$R_{AA}^{exp}$ and $R_{AA}^{th}$ are experimental and theoretical predictions for suppression factor, respectively, and $\sigma$ is related to experimental error.\\
We use the Minuit package \cite{Minuit.James:1975dr} for our parameter optimization process, which is a powerful tool that enables us to achieve high accuracy in minimizing chi-squared values.\\
We present our results for the nuclear modification factor in Fig.(\ref{fig: ALICE 2018}) to Fig.(\ref{fig: ATLAS 2022}) over a wide range of $P_T$ for all the three energy loss approaches. These results are compared with ATLAS 2022 data and ALICE data in 2018, 2021, and 2022. In addition, we repeat our fitting procedure for the momentum interval $2 <P_T<12$ (GeV), as shown in Fig.(\ref{fig: Intermediate $P_T$}), \textcolor{color}{to avoid significant errors of $R_{AA}$ data associated with small and large $P_T$s. We will then be able to reevaluate and compare our models within this specific range}. This range, known as the intermediate $P_T$ range, is of particular interest in the study of heavy-ion collisions as it encompasses a transition region between low $P_T$ and high $P_T$. This range of $P_T$ enables us to explore the interaction between high-energy scattering events and the collective properties of the QGP.\\
In tables (\ref{tab: ALICE 2018}) to (\ref{tab: ATLAS 2022}), we summarize the obtained results for the free parameters, $N$ and $K$, as well as the $\chi^2$ values for each model. Note that the charm quark distribution has been computed for $ P_T> 1$ GeV. Therefore, our results within $P_T< 1$ GeV region are invalid. To avoid errors resulting from unreliable regions, we calculate the $\chi^2$ value for data with $P_T > 1.5$ GeV. \\
As can be seen from the charts, in general, all three energy loss approaches agree well with experimental data from ALICE and ATLAS. Although they describe intermediate $P_T$ values better than large $P_T$s, there is a slight deviation from the experimental data observed for high $P_T$. That's because all models face divergence at the limits of integration and different approaches have used different methods to resolve this issue. It is anticipated that the introduction of novel methods capable of effectively resolving the divergence associated with the upper and lower limits of integration would yield improved outcomes for both small and large $P_T$ values.\\
\begin{figure}[htb]
	\begin{center}
		\vspace{0.4cm}
	\resizebox{0.65\textwidth}{!}{\includegraphics{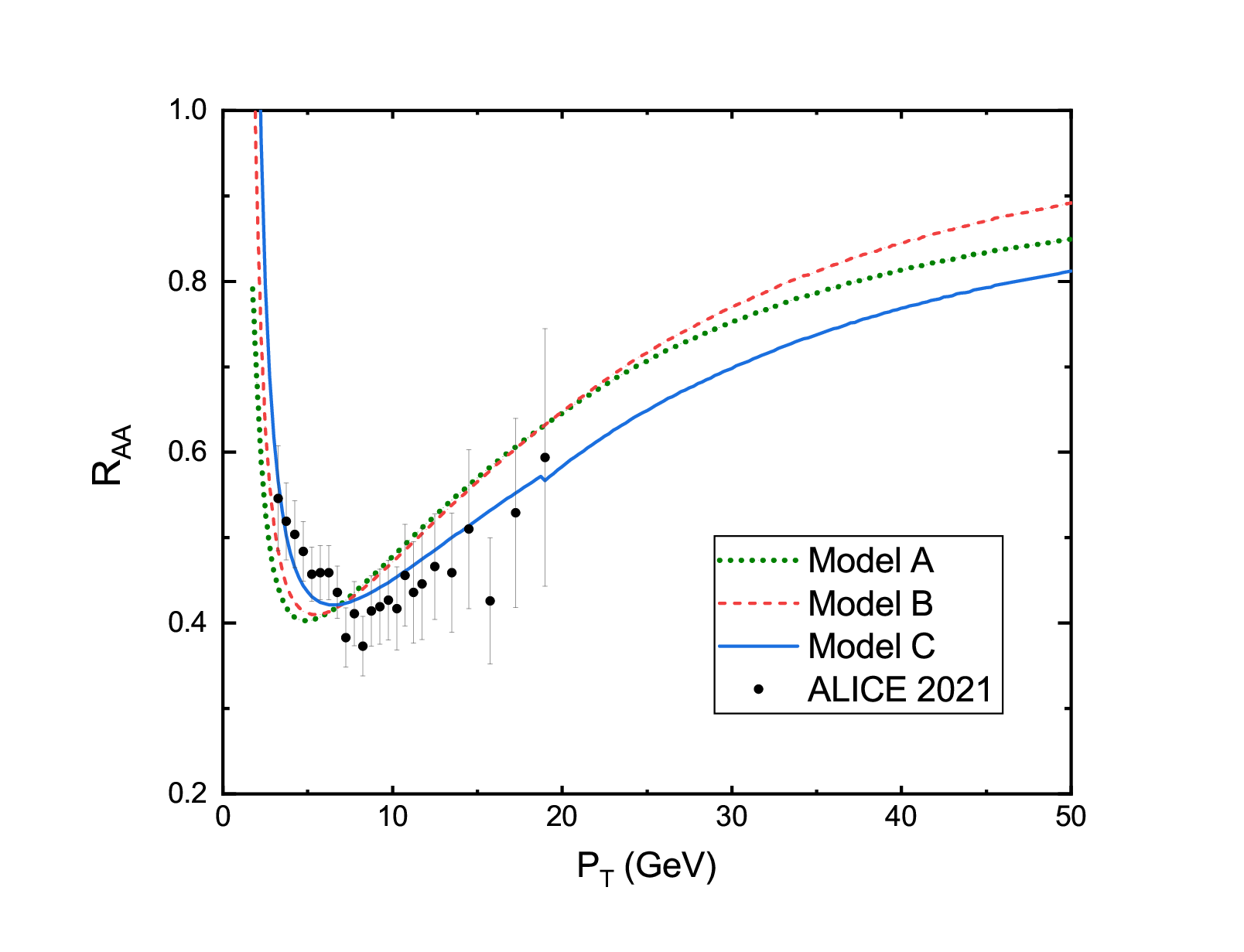}}
	\caption{ Nuclear modification factor of charm quark in the (20–40\%) centrality Pb-Pb collisions at $\sqrt{S_{NN}}=5.02$ TeV compared with ALICE 2021 data at forward rapidity ($2.5<y< 4$) \cite{ALICE:2020sjb}. }\label{fig: ALICE 2021}
\end{center}
\end{figure}
\begin{figure}[htb]
	\begin{center}
		\vspace{0.4cm}
	\resizebox{0.65\textwidth}{!}{\includegraphics{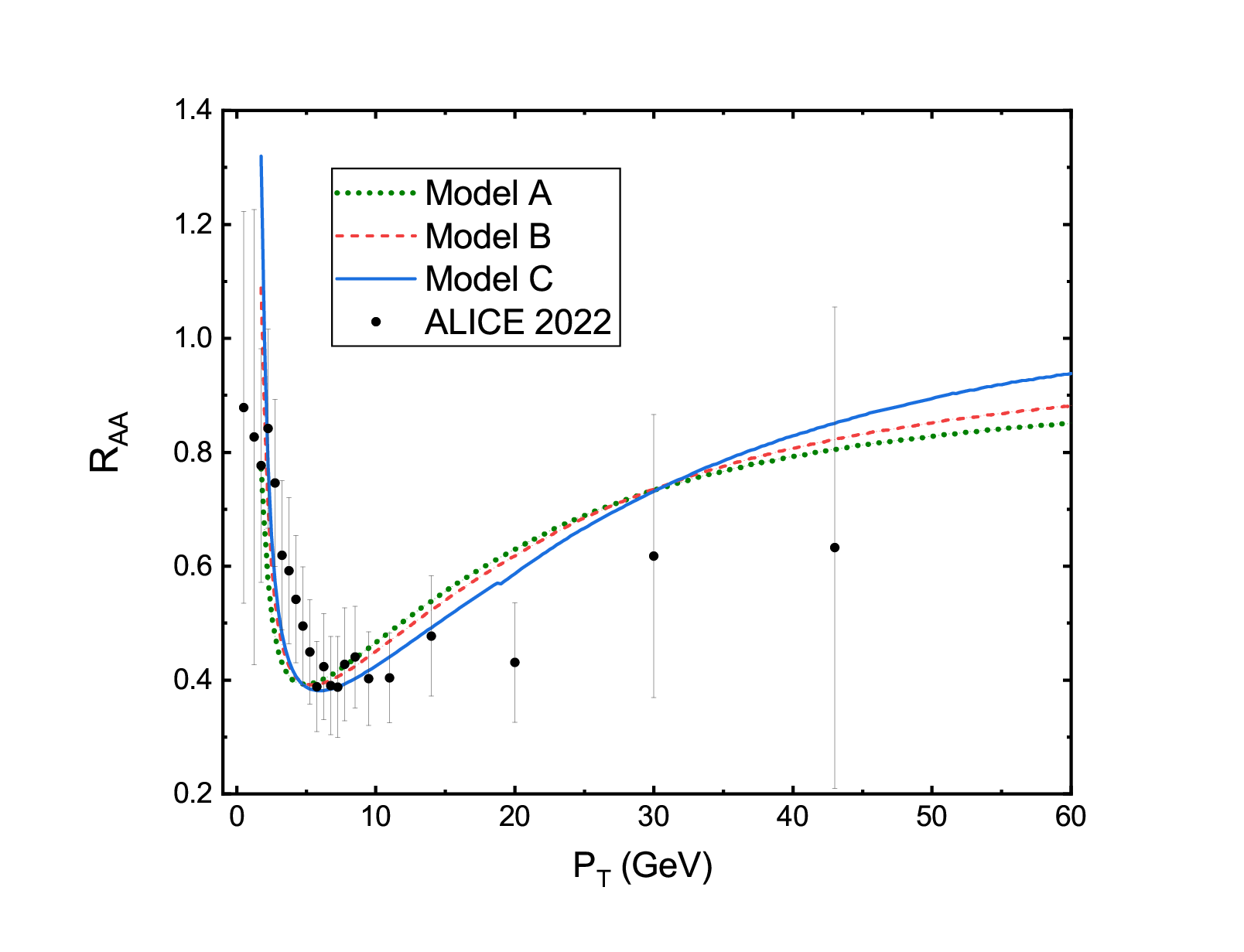}}
	\caption{ Nuclear modification factor of charm quark in the (30–50\%) centrality Pb-Pb collisions at $\sqrt{S_{NN}}=5.02$ TeV compared with ALICE 2022 data at mid-rapidity ($|y|<0.5$) \cite{ALICE:2021rxa}. }\label{fig: ALICE 2022}
\end{center}
\end{figure}
Figure (\ref{fig: ALICE 2018}) corresponds to the ALICE 2018 dataset, and Table (\ref{tab: ALICE 2018}) presents the fitting outcomes of these data for the three collisional energy loss models. As $\chi^2$ values show, for the 2018 dataset, there is no significant difference observed among the three approaches.  This lack of differentiation can be attributed to the limited number of data points available in 2018, coupled with their relatively high error margin (averaging 0.2). Hence, these data lack sufficient resolution to show the difference between the models and are not a good criterion for concluding.\\
In the ALICE 2021 and 2022 experiments, the number of data points has increased, and their errors have decreased. Consequently, these datasets provide a more reliable reference for investigating and comparing different energy loss models.
In the ALICE 2021 experiment, the measured data is related to muons decaying from D meson ($Pb+Pb \rightarrow{Muon + X}$), while in the ALICE 2022 experiment, the interaction is related to D meson production ($Pb+Pb \rightarrow{D0 + X}$). Consequently, the ALICE 2021 data has a smaller error compared to the ALICE 2022 data (the average error in the ALICE 2022 data is approximately 0.1, whereas the average error in the ALICE 2021 data is roughly half of that value).
However, ALICE's 2022 data cover a broader range of $P_T$ values. Therefore, they are more suitable for globally comparing energy loss models. Conversely, for the intermediate range of $P_T$ values, the 2021 data are a better reference due to their smaller error.\\
Figure (\ref{fig: ALICE 2021}) presents the results obtained for the ALICE 2021 data. As it is evident from Table (\ref{tab: ALICE 2021}) and the $\chi^2$ values, Model C performs better than other models in the range of intermediate $P_T$ values. This indicates that the HTL mechanism effectively describes the intermediate $P_T$ range, suggesting that Model C is a more suitable choice for this region. Furthermore, Model B outperforms Model A as expected.\\
Figure (\ref{fig: ALICE 2022}) illustrates the obtained results for the ALICE 2022 data. The ALICE 2022 data, covering a wider range of $P_T$ values, are more suitable for global analysis and comparison of different approaches. According to Table (\ref{tab: ALICE 2022}), for all $P_T$ ranges, Model B performs slightly better than other models. However, in general, there is no significant difference between the three energy loss models in the global analysis. We need more data points with higher $P_T$ values to make a definite conclusion. The fitting outcomes obtained from the ALICE 2022 dataset do not indicate a global advantage of Model C. Note that in our analysis, we have considered the range of $2 < P_T < 12 $ (GeV) as intermediate $P_T$s, while in Model C \cite{Braaten:1991we} the boundary between soft momentum and hard momentum occurs around $P_T =$ \SI{20}{\giga\electronvolt}. Consequently, we should note that in the review of Model C, most of the available data are evaluated using the HTL mechanism. To obtain better results for a global investigation of Model C, the boundary for these two regions may need to be modified at higher center-of-mass energies.\\
Finally, the ATLAS 2022 data were analyzed to compare with the ALICE data; Figure (\ref{fig: ATLAS 2022}). However, the limited number of ATLAS data points and their higher error compared to the ALICE 2021 data make it difficult to differentiate between the three methods of energy loss. Nonetheless, as seen in Table (\ref{tab: ATLAS 2022}), in this case as well, all the models provide better descriptions of the data in the range of intermediate $P_T$ values in comparison by the whole $P_T$ region.\\


\begin{table}[hb]
	\centering
	\caption{The best fitting results for different approaches for ALICE 2018 data}
	\label{tab: ALICE 2018}
	\begin{tabular}{|c|c|c|c|c|}
		\hline
		\textbf{Approach} & \hspace{3mm} Range of $P_T$ (GeV) \hspace{3mm} & \textbf{$\chi^2$} & \textbf{N} & \textbf{k} \\
		\hline
		\multirow{2}{*}{Model A}  &  All $P_T$ & \hspace{1cm} 0.86 \hspace{1cm} & \hspace{1cm} 0.74 \hspace{1cm} & \hspace{1cm} 1.2 \hspace{1cm} \\ 
         & $2<P_T<12$ & 0.32 & 0.75 & 1.01 \\
		\hline
		\multirow{2}{*}{Model B}  & All $P_T$ & 0.81 & 0.75 & 1 \\ 
         & $2<P_T<12$ & 0.35 & 0.72 & 1 \\
		\hline
		\multirow{2}{*}{Model C}  & All $P_T$ & 0.82 & 0.72 & 1.4 \\ 
         & $2<P_T<12$ & 0.40 & 0.50 & 2.3 \\
		\hline
	\end{tabular}
\end{table}

\begin{table}[h]
	\centering
	\caption{The best fitting results for different approaches for ALICE 2021 data}
	\label{tab: ALICE 2021}
	\begin{tabular}{|c|c|c|c|c|}
		\hline
		\textbf{Approach} & \hspace{3mm} Range of $P_T$ (GeV) \hspace{3mm} & \textbf{$\chi^2$} & \textbf{N} & \textbf{k} \\
		\hline
		\multirow{2}{*}{Model A}  &  All $P_T$ & \hspace{1cm} 2.2 \hspace{1cm} & \hspace{1cm} 1.15 \hspace{1cm} & \hspace{1cm} 1 \hspace{1cm} \\ 
         & $2<P_T<12$ & 2.5 & 1.08 & 1.1 \\
		\hline
		\multirow{2}{*}{Model B}  & All $P_T$ & 1.7 & 1.06 & 1 \\ 
         & $2<P_T<12$ & 1.9 & 1.01 & 1.1 \\
		\hline
		\multirow{2}{*}{Model C}  & All $P_T$ & 0.67 & 1.16 & 1 \\ 
         & $2<P_T<12$ & 0.76 & 1.15 & 1 \\
		\hline
	\end{tabular}
\end{table}

\begin{table}[h]
	\centering
	\caption{The best fitting results for different approaches for ALICE 2022 data}
	\label{tab: ALICE 2022}
	\begin{tabular}{|c|c|c|c|c|}
		\hline
		\textbf{Approach} & \hspace{3mm} Range of $P_T$ (GeV) \hspace{3mm} & \textbf{$\chi^2$} & \textbf{N} & \textbf{k} \\
		\hline
		\multirow{2}{*}{Model A}  &  All $P_T$ & \hspace{1cm} 1.06 \hspace{1cm} & \hspace{1cm} 1.18 \hspace{1cm} & \hspace{1cm} 1 \hspace{1cm} \\ 
         & $2<P_T<12$ & 0.73 & 0.55 & 3.3 \\
		\hline
		\multirow{2}{*}{Model B}  & All $P_T$ & 0.88 & 1.11 & 1 \\ 
         & $2<P_T<12$ & 0.52 & 1 & 1 \\
		\hline
		\multirow{2}{*}{Model C}  & All $P_T$ & 0.97 & 0.98 & 1.63 \\ 
         & $2<P_T<12$ & 0.24 & 1.2 & 1.08 \\
		\hline
	\end{tabular}
\end{table}

\begin{table}[h]
	\centering
	\caption{The best fitting results for different approaches for ATLAS 2022 data}
	\label{tab: ATLAS 2022}
	\begin{tabular}{|c|c|c|c|c|}
		\hline
		\textbf{Approach} & \hspace{3mm} Range of $P_T$ (GeV) \hspace{3mm} & \textbf{$\chi^2$} & \textbf{N} & \textbf{k} \\
		\hline
		\multirow{2}{*}{Model A}  &  All $P_T$ & \hspace{1cm} 0.25 \hspace{1cm} & \hspace{1cm} 0.85 \hspace{1cm} & \hspace{1cm} 1 \hspace{1cm} \\ 
         & $2<P_T<12$ & 0.11 & 0.74 & 1.22 \\
		\hline
		\multirow{2}{*}{Model B}  & All $P_T$ & 0.24 & 0.78 & 1 \\ 
         & $2<P_T<12$ & 0.12 & 0.70 & 1.19 \\
		\hline
		\multirow{2}{*}{Model C}  & All $P_T$ & 0.19 & 0.8 & 1.07 \\ 
         & $2<P_T<12$ & 0.17 & 0.66 & 1.56 \\
		\hline
	\end{tabular}
\end{table}

\FloatBarrier
\section{ Summary and Conclusions} \label{sec: conclusion}

Our study employed the Fokker-Planck equation to investigate the evolution of transverse momentum distribution functions of charm quarks produced in lead-lead collisions at 5.02 TeV.
During the evolution, we considered three different approaches for collisional energy loss to compare these approaches with each other. Our purpose was to assess their compatibility with the latest experimental data. 
We have found that the recent data have sufficient precision to distinguish among these different models, particularly in the region of intermediate transverse momentum. Although published data older than 2018, do not exhibit significant differences between the various energy loss models, mainly due to their limited quantity and large errors.
In general, all three energy loss approaches describe the range of intermediate $P_T$ better than small or large $P_T$ regions. Among these models, the model proposed by Braaten and Thoma \cite{Braaten:1991we} provides a better description of the intermediate $P_T$ range, in comparison with the other energy dissipation methods. Indicating that the HTL mechanism is an appropriate mechanism for average $P_T$s.
For the global analysis, we considered ALICE 2022 data as a benchmark, because it covers a wider region of $P_T$s. There was no significant difference between the $\chi^2$ values for different energy loss models.
In fact, the major difference between these models lies in managing the convergence at high and low momenta and the method of field regularization. Therefore, by increasing the number of data points for small and large $P_T$s, and decreasing their error, we should be able to distinguish between these energy loss models more effectively for global analysis. Also, It is expected that such evolution will be employed in the near future for the b quark distribution function. At present, the available data on the b quark distribution function are insufficient to determine the validity of a proper model.\\

\section*{Acknowledgments}
Special thanks go to Dr. Samira Shoeibi for providing guidance in using the Minuit package. This work is supported by the Ferdowsi University of Mashhad under
grant numbers 3/58322 (1401/07/23).

\clearpage
\section*{Data Availability Statement}\label{data} This manuscript has associated data in a data repository. [Authors’ comment: The data sets generated during the current study
are available from the corresponding author on reasonable request.]

\appendix
\renewcommand{\thefigure} {\arabic{figure}}
\setcounter{figure}{0}
\section*{Appendix}\label{sec:appendix}

\textbf{A. Energy loss } As mentioned before, in order to calculate drag and diffusion coefficients in the Fokker-Planck equation, we must calculate the energy loss of heavy quarks while passing through the plasma and consider both modes of energy loss through collisions and radiation. Here, we introduce three common approaches for calculating collisional energy loss and one of the most common approaches to calculating radiant energy loss. \\
The first calculation of collisional energy loss (Model A in our article) is proposed by Bjorken \cite{Bjorken:1982tu} which is : 
\begin{equation}
	-\frac{dE}{dx}=\frac{16\pi}{9}\alpha_s^2T^2\ln\left(\frac{4pT}{k_D^2}\right)\left[\exp\left(-\frac{k_D}{T}\right)\left(1+\frac{k_D}{T}\right)\right]
\end{equation}
$P$ is the momentum of the particle, $T$ is the temperature of the plasma, and $k_D = \sqrt{3} m_g$.  We also have:
\begin{equation}
m_g^2 = \frac{4\pi\alpha_s T^2}{3}(1+\frac{n_f}{6}) 
\end{equation}
\\
Another approach for calculating collisional energy loss is presented by Thoma and Gyulassy  \cite{Thoma:1990fm}. Through this approach which is our second model, we have:
\begin{equation}
-\frac{dE}{dx} = \frac{16\pi}{9} \alpha_s^2 T^2 \ln\left(\frac{k_{\text{max}}}{k_D}\right) \frac{1}{\nu^2}\left[\nu + \frac{\left(\nu^2 - 1\right)}{2} \ln\left(\frac{1+\nu}{1-\nu}\right)\right]
\end{equation}
In which:
\begin{equation}
k_{\text{max}} \approx \frac{4pT}{\sqrt{p^2 + M^2} - p + 4T}
\end{equation}
\\
Model C for collisional energy loss \cite{Braaten:1991we} involved calculating the energy loss of a quark with energy E in two different limits: $E \ll \frac{M^2}{T} \text{ and } E \gg \frac{M^2}{T}$ \\
A QED calculation has been used to determine contributions to the energy loss for some parts of the calculation. To achieve this, "e" in the QED calculations will be replaced by the $g_s=\frac{4}{3}\sqrt{4\pi\alpha_s}$ in the QCD calculations. The thermal photon mass $m=eT/3$ is also replaced by the thermal gluon mass which is $m_g=g_s T \sqrt{\frac{1+n_f/6}{3}}$\\
So for the $E \ll \frac{M^2}{T}$ limit we will have: 
\begin{equation}
	-\frac{dE}{dx} = \frac{8\pi\alpha_s^2 T^2}{3}(1+\frac{n_f}{6})\left[\frac{1}{v}-\frac{1-v^2}{2v^2}\ln\left(\frac{1+v}{1-v}\right)\right]\ln\left(\frac{2^{n_f/(6+n_f)}B(v)ET}{m_g M}\right)	
\end{equation}
$B(v)$ is a smooth function that starts at $B(0)=0.604$, increases to $B(0.88)=0.731$, and then decreases to $B(1)=0.629$.\\
And in the $E \gg \frac{M^2}{T}$  limit, we have:\\
\begin{equation}
	-\frac{dE}{dx} = \frac{8\pi\alpha_s^2 T^2}{3}(1+\frac{n_f}{6})\ln\left(2^{\frac{n_f}{12+2n_f}} 0.920 \frac{\sqrt{ET}}{m_g}\right)
\end{equation}
A smooth connection between two limits is required for the intermediate region, E $\approx M^2/T$. Calculations indicate that we can use the first equation up to $E_{cross} = 1.8 M^2/T$ and then switch to the second one.\\
\\
Also, the radiative energy loss of a heavy quark in a QGP is calculated as follows:
\begin{equation}
	-\frac{dE}{dx} = 24\alpha_s^3 \rho_{\text{QGP}} \frac{1}{\mu_g} (1-\beta_1) \left(\sqrt{\frac{1}{1-\beta_1} \ln\frac{1}{\beta_1}} - 1\right) F(\delta)
\end{equation}
\begin{equation}
	F(\delta) = 2\delta - \frac{1}{2} \ln\left(\frac{1+\frac{M^2}{s}e^{2\delta}}{1+\frac{M^2}{s} e^{-2\delta}}\right) - \left(\frac{\frac{M^2}{s}\sinh(2\delta)}{1+2\frac{M^2}{s}\cosh(2\delta)+\frac{M^4}{s^2}}\right) 
\end{equation}
\begin{equation}
    \delta = \frac{1}{2} \ln \left[ \frac{1}{1 - \beta_1} \ln \left( \frac{1}{\beta_1} \right) \left( 1 + \sqrt{1 - \frac{1 - \beta_1}{\ln (1/\beta_1)}} \right)^2 \right]
\end{equation}
\begin{equation}
     C = \frac{3}{2} - \frac{M^2}{48E^2 T^2 \beta_0} \ln \left[ \frac{M^2+6ET(1+\beta_0)}{M^2+6ET(1-\beta_0)} \right]
\end{equation}
for more details see \cite{Saraswat:2017vuy}.\\
\

\begin{figure}[h]
	\begin{center}
		\vspace{0.4cm}
	\resizebox{0.45\textwidth}{!}{\includegraphics{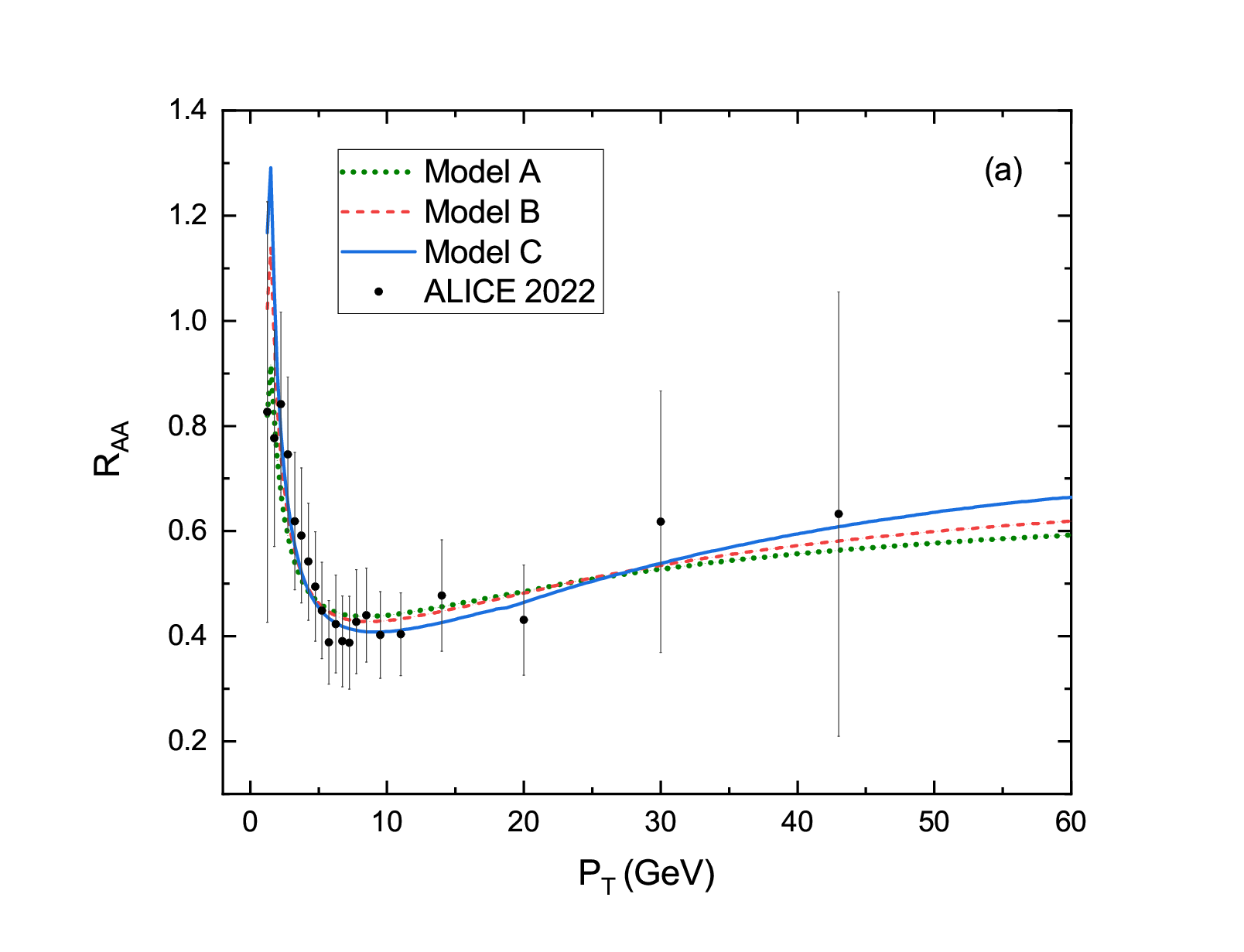}}
        \resizebox{0.45\textwidth}{!}{\includegraphics{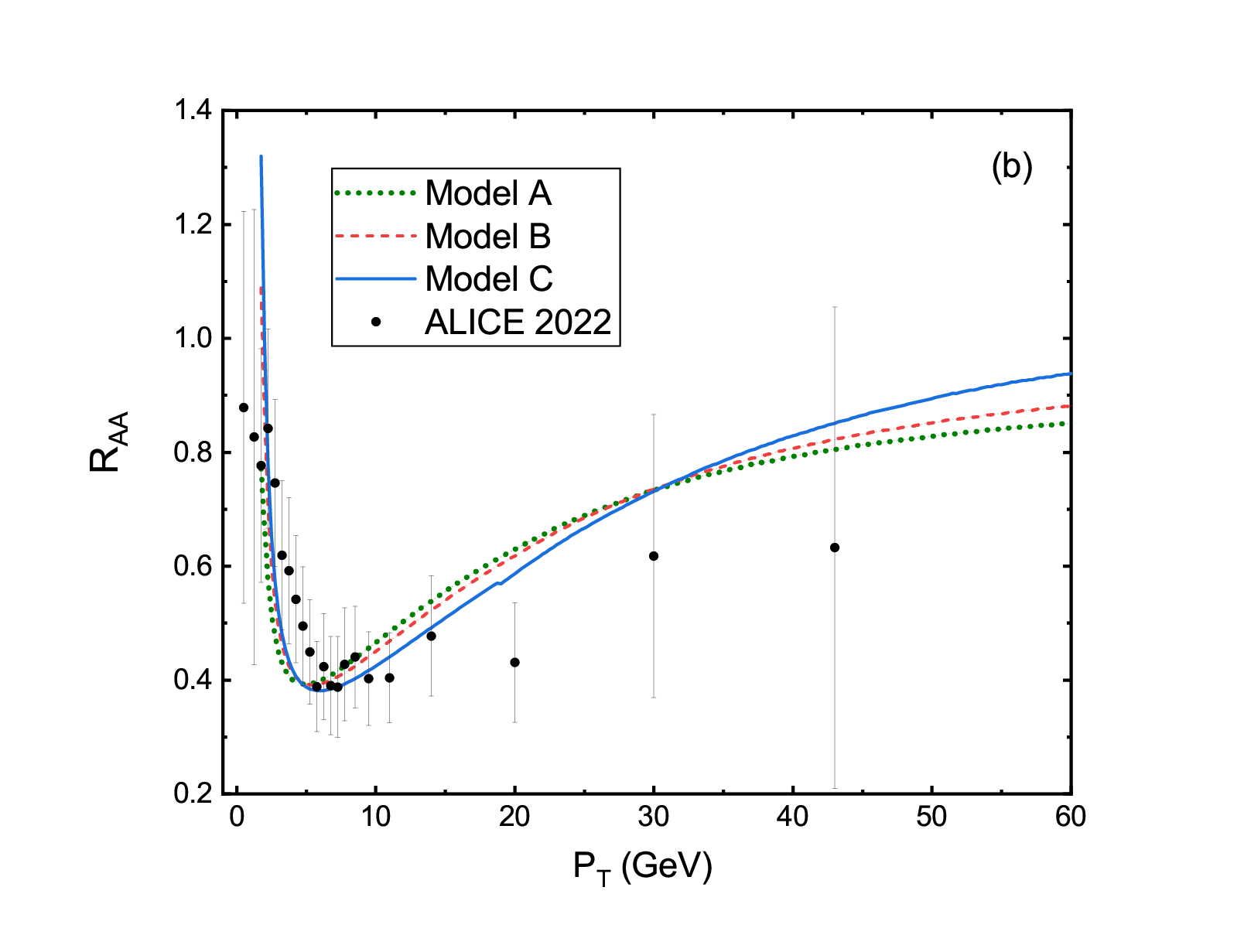}}

	\caption{ Left panel (a) compares three energy loss models through $R_{AA}$ of D meson and
 right panel (b) compares these models through $R_{AA}$ of charm quark. }\label{fig: compare}
  \end{center}
\end{figure}

\textbf{B. Hadronization } In order to find the $P_T$ distribution function for D meson, one can use the Peterson fragmentation function which is $D_c^D (z) = \frac{1}{{z\left(z - \frac{1}{z} + \frac{\epsilon}{{1-z}}\right)^2}}$ \cite{Peterson:1982ak}. Here, $z = \frac{P_D}{P_C}$ is the momentum fraction of the D meson which is fragmented from the charm quark.\\
We extend our calculations up to the hadronization stage for the ALICE 2022 dataset to assess the effect of hadronization on $R_{AA}$ shape.
Figure (\ref{fig: compare}) compares energy loss models before and after hadronization.\\

\
It can be seen from Fig. (\ref{fig: compare}) that the performance of the three models, in comparison to each other, remains relatively unchanged before and after hadronization. Therefore, the comparison of changes in dE/dx appears to be valid up to the pre-hadronization stage, aligning with the conventional assumption in other studies \cite{Tripathy:2017kwb, Tripathy:2016hlg, Qiao:2020yry, vanHees:2005wb}.\\

\

\begin{figure}[ht]
	\begin{center}
		\vspace{0.3cm}
	\resizebox{0.5\textwidth}{!}{\includegraphics{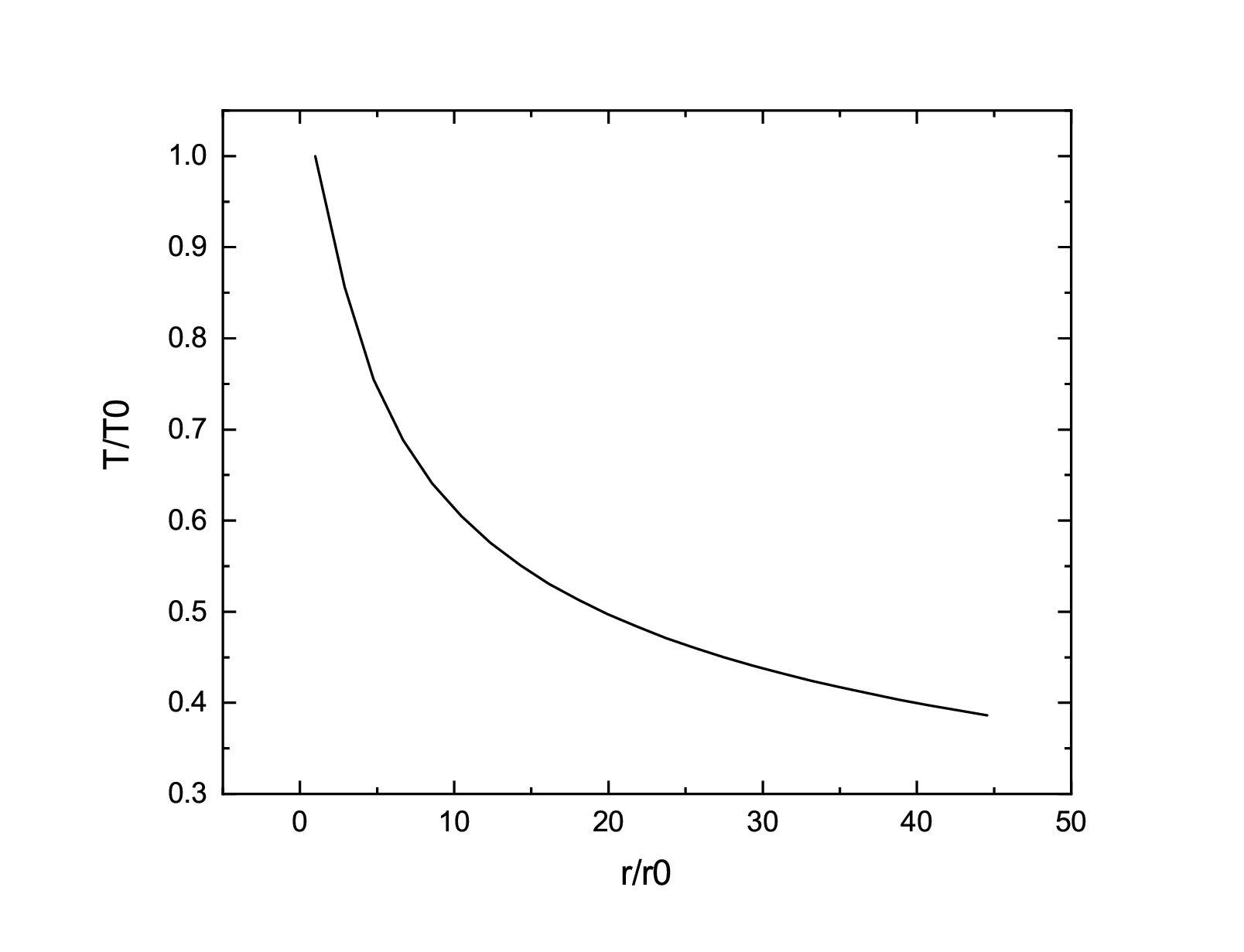}}
	\caption{The profiles of temperature}\label{fig: temp.profile}
\end{center}
\end{figure}

\textbf{C. Hydrodynamic evolution equations of QGP } To consider the profile of temperature, it is important to note that we have solved the evolution equation using the Bjorken flow. Figure (\ref{fig: temp.profile}) shows the temperature profile resulting from equation (\ref{Eq:eq1}).\\
We have constructed the hydrodynamic evolution equations of the QGP in the Milne coordinates ($\tau$,$r$,$\phi$,$\eta$) as Bjorken flow, where:\\
\begin{equation}
\begin{aligned}
\tau = \sqrt{t^2 - z^2}, & \quad
\eta = \tanh^{-1}\left(\frac{z}{t}\right), \\
r = \sqrt{x^2 + y^2}, & \quad
\phi = \tan^{-1}\left(\frac{y}{x}\right).
\end{aligned}
\end{equation}

in which r and $\phi$ express the transverse plane and $\eta$ is the rapidity (along the beam direction z).\\
According to the boost-invariance along the $\eta$, rotational and translational invariance in the transverse plane, as well as the reflection symmetry under $\eta \leftrightarrow -\eta$, the only flow due to these symmetries is $u^\mu = (u^\tau, u^x, u^y, u^\eta) = (1, 0, 0, 0)$. This means that ($r$, $\phi$, $\eta$) are independent of the macroscopic physical quantities.\\
Considering dissipative hydrodynamics, the energy-momentum tensor can be defined as:
\begin{equation}
    T^{\mu\nu} = \langle p^\mu p^\nu \rangle = \epsilon u^\mu u^\nu + P \Delta^{\mu\nu} + \pi^{\mu\nu}
\end{equation}
where $\epsilon$ and $P$ are energy density and pressure, which are functions of the QGP temperature \cite{Chattopadhyay:2018apf, Grozdanov:2015kqa}. Also we have $\Delta^{\mu\nu} = g^{\mu\nu} + u^\mu u^\nu$  and $\mu^{\mu\nu}$ is the shear stress tensor.\\
The evolution equations for the $\epsilon$ and $u^\mu$ are extracted through the $\partial_\mu T^{\mu\nu} = 0$ as follows:
\begin{equation}
    \dot{\epsilon} + (\epsilon + P)\theta + \pi^{\mu\nu} \sigma_{\mu\nu} = 0
\end{equation}
\begin{equation}
    (\epsilon + P) \dot{u}^\alpha + \nabla^\alpha P + \Delta_\nu^\alpha  \partial_\mu  \pi^{\mu\nu} = 0
\end{equation}
There are different expressions for the shear stress tensor. We have used the published expansion of the $\pi^{\mu\nu}$ up to the third-order terms \cite{Chattopadhyay:2018apf, Jaiswal:2013vta, Chattopadhyay:2014lya}.\\
The time evolution of $\epsilon$ and $\pi$ are as follows:
\begin{equation}
    \frac{d\epsilon}{d\tau} = -\frac{1}{\tau}\left(\epsilon + P - \pi\right)
\end{equation}
\begin{equation}
    \frac{d\pi}{d\tau} = -\frac{\pi}{\tau_\pi} + \frac{1}{\tau}\left(\frac{4}{3}\beta_\pi - \lambda\pi - \chi \frac{\pi^2}{\beta_\pi}\right)
\end{equation}
where $\beta_\pi = \frac{4P}{5}$, $\lambda = \frac{38}{21}$ and $\chi = \frac{72}{245}$ are third-order contribution coefficients of shear stress tensor expansion if we take $\epsilon = 3 P $.\\

\clearpage

\end{document}